\documentclass[conference]{IEEEtran}
\IEEEoverridecommandlockouts
\usepackage{cite}
\usepackage{amsmath,amssymb,amsfonts}
\usepackage{algorithmic}
\usepackage{graphicx}
\usepackage{textcomp}
\usepackage{xcolor}
\def\BibTeX{{\rm B\kern-.05em{\sc i\kern-.025em b}\kern-.08em
    T\kern-.1667em\lower.7ex\hbox{E}\kern-.125emX}}
\begin{document}

\title{Implementation of Honeynet and Honeypot in Network Infrastructure in Production Network}

\author{\IEEEauthorblockN{ Nawshad Ahmed Evan}
\IEEEauthorblockA{\textit{School of Computing and Creative Technologies} \\
\textit{University of the West of England}\\
Bristol BS16 1QY, United Kingdom \\
nawshadevan@gmail.com}
\and
\IEEEauthorblockN{ Md Raihan Uddin}
\IEEEauthorblockA{\textit{Department of Electrical and Computer Engineering} \\
\textit{The University of Alabama in Huntsville}\\
Huntsville, AL 35899, USA \\
mu0016@uah.edu}
}

\maketitle

\begin{abstract}
Network infrastructure in a production environment is increasingly targeted by attackers every day. Many resources and services now rely on the internet, making network infrastructure one of the most critical parts to protect, as it hosts numerous company resources and services. Several solutions have already been proposed to prevent attacks, minimize damage, and divert hackers and intruders. Among these, the honeypot stands out as a highly effective tool; it is designed to mimic both a scanner and an attacker, diverting and misleading them within a simulated, production-level environment. This paper will demonstrate the use of a honeynet where a honeypot acts like a real resource to deceive the attacker and analyze their behavior.
\end{abstract}


\section{Introduction}
Security is a top priority in both social and corporate life. To maintain strong security, it is essential to understand data protection techniques and perform regular security checks \cite{rathore2017social}. The world is becoming more challenging in terms of security, making a safe environment a primary focus for individuals and companies. As technology advances, each new day brings fresh security challenges \cite{lin2018survey}.

In a production environment, network security is a significant challenge. Protecting the network is one of the most important factors, whether it's operating over the internet, a LAN, or other methods, and this is true for any size of business. While no system is completely immune to attacks, a strong and effective security system is essential for safeguarding customer data. A good security system helps businesses reduce the risk of data theft and sabotage \cite{sagduyu2019iot}. Digitally connected devices and applications are becoming a part of every aspect of our lives—homes, offices, cars, and even our bodies. All of these gadgets are becoming "smarter" to take advantage of being connected to the internet.

The Internet of Things (IoT) is growing rapidly, which has expanded the attack surface far beyond traditional enterprise IT infrastructures. It is important to understand the security risks of IoT before discussing its features \cite{mozhaev2017multiservice}. This study explains how Honeypots and Honeynets can be used to increase the security of a network's infrastructure.

Honeypots are designed to improve security by detecting unauthorized attempts to access a data system. Modern network security often includes firewalls, intrusion detection systems, and encryption. However, today's environment requires more proactive methods to identify, redirect, and contain unwanted access. Honeypots offer a proactive approach to these network security concerns \cite{matin2019malware}. Honeypots are typically hosted on servers that simulate different environments to appear like a real network. An attacker can enter the honeynet from any machine, which allows the system to monitor the attacker's behavior and divert threats away from the real network. A honeynet simulates a genuine network more accurately than a single honeypot, making it suitable for large or complex environments where attackers might be drawn to what seems like an authentic and appealing target \cite{ren2021theoretical}.

All network infrastructure resources—including hardware, software, systems, and devices—work together to enable connectivity, control, and communication. This ranges from servers to Wi-Fi routers. This infrastructure also includes the network protocols that allow users and systems to communicate and interact \cite{fox2019deployment}.

This study will propose a secure, hybrid-designed model for network infrastructure security to provide trusted communication and reliable service. The main challenge is to integrate honeypots and honeynets with firewalls and analyze hacker footprints to improve security. We will use two types of honeypots: one will be placed in front of the honeynet's production router to act as a fake web server, tricking attackers into spending their time there. The other will be placed behind the router, where it can analyze attacker behavior in a controlled environment. The honeynet is set up before the main network infrastructure to ensure security and continuous service in a production environment.

\section{Related Works}
New methods and technologies are needed for network security and forensics. Honeypots are widely used for network security, as these tools can track intruders and monitor hacker activity. Modern communication has connected the world and made the internet accessible to everyone. To maintain security, it is crucial to continuously develop new strategies and upgrade firewalls and intrusion detection systems \cite{sheikh2020network}. Honeypot security solutions are constantly being improved and used in new ways. Even traditional decoy honeypots have many uses for security engineers and can sometimes help predict security vulnerabilities.

Cloud services and their user bases are growing rapidly. Companies are now more comfortable hosting their servers in the cloud to serve their clients. While internet and app services have made our lives easier, this expansion also increases vulnerability. To protect these services, Honeypot and Honeynet technologies must be improved by closely monitoring and analyzing attacker activity \cite{baykara2018novel}.

This paper will implement a technique to create a Honeynet and Honeypot. This method is designed to deceive hackers with fake content, diverting them from the real network and potentially protecting it. However, if a hacker carefully examines the material and detects the deception, they might erase their traces before we can collect them and could even plan a larger attack. Poorly maintained or misused honeypots could also increase risks from hackers \cite{krishnaveni2018survey}.

Understanding hacker attack patterns is essential for network engineers and researchers. Engineers regularly review security reports and logs to measure these patterns, which is a key part of developing and maintaining security. By comparing an attacker's habits and attack stages, researchers have noticed that attacks on weekdays can have different outcomes than those on weekends. Although some days stand out, attack patterns can vary. By studying these patterns, experts can better understand an attacker's behavior and adjust the network's security architecture to defend against it. Hackers can attack from outside the network or by compromising an internal user's computer. Honeypots and Honeynets help security analysts by recording both external and internal threats and vulnerabilities \cite{leech2024ten}.

An employee's lack of security awareness can make a company vulnerable. Employees with low awareness might accidentally reveal company secrets. Therefore, security awareness training is essential for network security. Many companies are now training their employees on security to help them manage risks and vulnerabilities. This training often focuses on basic communication, technical skills, and how to recognize threats like phishing and ransomware \cite{malecki2019best}.

Hackers attack individuals and corporations to steal data, often aiming to steal money or hold data for ransom in exchange for money or cryptocurrencies. Financial firms and organizations that hold public data are major targets, but this does not mean small businesses are immune. Hackers will also attack any network that appears insecure. As a result, security must be a top priority for any organization, regardless of its size.

In security research, a honeypot is a valuable tool for carefully examining hacker activity. It allows security experts to see how an attacker penetrates a system, elevates their privileges, and moves through the network. Security organizations, research institutions, and government agencies are all assessing these threats to find solutions. With a honeypot, an attacker's actions can be examined in a secure environment, either physical or virtual, preventing a breach of the actual network \cite{kyriakou2018container}.

\section{Proposed Approach}
This section explains our methodology. We will introduce the topic, present the data we collected, analyze it, and then validate our findings to show the project's reliability. The goal of this project is to implement a solution for real-time threats to network infrastructure. The analysis and project discussion will be illustrated with a diagram to clarify our method and structure.

This paper will create a Honeypot within a Honeynet. A honeynet is a network where all devices are decoys but are designed to function like real ones. A honeypot is a single fake server or workstation that imitates a real resource \cite{fan2019honeydoc}. Active protection is a relatively new concept in information technology. While some have tried to define this term, many definitions are incomplete or miss key features of this security approach. Some have blended active and offensive security strategies.

This study examines the features of active protection techniques like Honeypots and Honeynets, including their benefits and limitations \cite{wu2019active}. Honeypot systems have been around for over a decade and have improved and adapted over time, thanks to programs like The Honeynet Project and Project Honeypot.

Despite their benefits, honeypots are not commonly used in businesses. This may be due to the challenges of installing and managing a honeypot, as well as a lack of understanding of its advantages. Compared to traditional firewalls, honeypots and honeynets offer advanced security features and provide valuable support for security engineers \cite{dowling2017zigbee}.

\subsection{Network Security Tactics and Types}
The foundation of network security management is the CIA triad: Confidentiality, Integrity, and Availability. Many network security tools are ineffective without a good strategy and proper administration. To manage a network without interruption and meet user needs, it's important to handle fault management, configuration, and performance analysis of the security infrastructure \cite{aminzade2018confidentiality}.

\textbf{Confidentiality} protects sensitive data from unauthorized users and attackers. Access to this data should be limited to authorized individuals. \textbf{Integrity} ensures that data is not altered by unauthorized people, maintaining its accuracy for providers. \textbf{Availability} ensures that authorized users can access the data when they need it. Any interruption that prevents users from accessing the service is considered a security violation. The service must be secure and consistently accessible to the right users \cite{yin2020hierarchically}.

Security methods can be categorized based on the component level: hardware, software, and cloud. \textbf{Hardware} is the physical part of a network or connected device. It is necessary to safeguard hardware from physical and digital attacks, as a network attack can damage hardware by overusing its resources and shortening its lifespan. \textbf{Software} is the graphical interface that provides services. A direct software infection can damage the hardware or the entire network, so software security must always be kept up-to-date. \textbf{Cloud} services are hosted on the internet and can be used from anywhere. It's crucial to protect this blended hardware and software environment in data centers using the CIA approach \cite{atieh2021assuring}.

\subsection{Conceptual Framework}
Network security engineers are always focused on preventive measures to protect their network. Security surveillance is needed to prevent network breakdowns, unauthorized access, and malicious users. Securing a commercial network often requires expensive hardware and software. A security framework for a network is typically based on its core design \cite{saigushev2018information}.

Since this project uses open-source components, its security approach was designed to be adaptable to any network topology. Engineers choose hardware based on their network design. This article discusses our specific network design and compares it to standard and hybrid security models.

We will use the GNS3 open-source simulator to build a decoy network topology where some devices will behave like real ones. We will configure network protocols and connect them to the internet with a real routing engine to make the simulation more realistic. This network will be placed next to the main production network's border router. A Windows-based Honeypot server will be configured behind the Honeynet router to analyze attacker activities and store logs. This honeypot will have defense and traceback systems to capture attacker footprints, trace them back to their source, and analyze them.

Tracing the attacker's IP will help reveal their location, hop count, and route path. The connection state of each hop will be presented in milliseconds (ms) for measurement. We will also check the source IP's domain information for identification \cite{nur2018record}.

\begin{figure}[ht]
    \centering
    \includegraphics[width=0.88\linewidth]{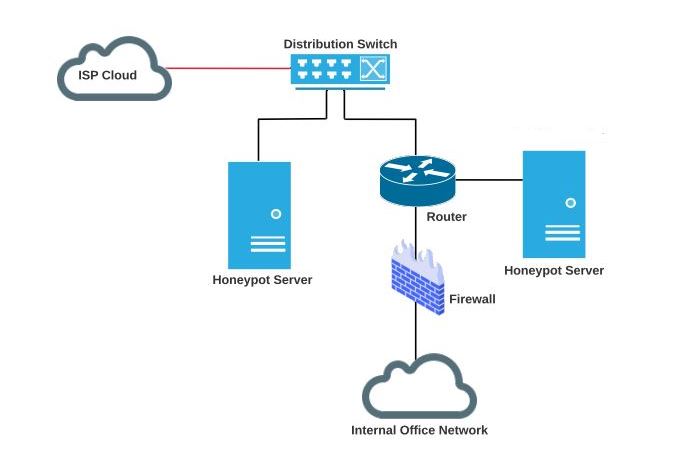}
    \caption{Hybrid network security infrastructure.}
    \label{fig:hybrid_diagram}
\end{figure}

An attacker will first try to find and enter the production network. To counter this, another honeypot will be set up in front of the Honeynet network router to deceive the attacker. It will present a fake message to make it seem like a legitimate web server. The goal is to make the attacker think they are in a real service environment, which will distract them on the fake web server. This decoy device is placed before the Honeynet router to entice hackers. However, if poorly designed, attracting hackers into a fake network can be dangerous. A comprehensive Honeynet is therefore designed to prevent hackers from gaining access to more devices if they become more aggressive \cite{krishnaveni2018survey}.

Next, an attacker machine will be placed in the network topology to generate random attacks using Kali Linux tools to evaluate the system. The logs will then be analyzed to understand the attacker's activity. This project offers a design that traps attackers, making it difficult for them to distinguish between real and fake networks. By searching the honeypot for resources, the attacker wastes time and leaves behind valuable footprints and logs, which helps us understand their attack patterns. This can reduce vulnerabilities and protect the real network infrastructure. The second honeypot will capture the attacker's source to determine their network origin and system details, identifying any other potential vulnerabilities.

\subsection{Comparison Between Regular and Hybrid Design}
Cyber attacks are one of the biggest criminal risks today, and security engineers are constantly taking steps to prevent them. Security breaches often happen due to a lack of awareness or insufficient precautions. Security measures rely on hardware, software, and cloud components to be effective \cite{tirumala2019survey}.

Different network security models use different tools. Most models include a firewall or a similar security device. Many security engineers believe a firewall is the most cost-effective solution for network security. A firewall offers services like intrusion detection and prevention, threat analysis, security logs, and more. The main goal of a firewall is to protect a network and provide uninterrupted service \cite{jingyao2019securing}.

However, a firewall alone cannot protect a network with critical services. Attackers today can bypass firewalls using methods like SQL injection, social engineering, application vulnerabilities, and exploiting IoT devices or employee ignorance. Once a firewall is breached, hackers can quickly infiltrate networks and access resources \cite{wildenauer2019hacking}.

\begin{figure}[ht]
    \centering
    \includegraphics[width=0.99\linewidth]{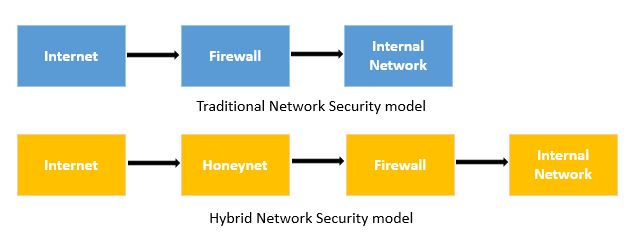}
    \caption{Comparison of traditional and hybrid secure network designs.}
    \label{fig:comparison_diagram}
\end{figure}

In addition to a firewall, other security control methods are needed. While a firewall is a crucial component, this traditional approach is well-known to everyone, including hackers.

The network model developed here uses a different security approach. As shown in the diagram, it features a three-layer security system: a decoy device that misleads the attacker, a fake network that manipulates them, and another decoy device that analyzes and tracks them. These three layers are placed before the firewall that protects the production network. This unique approach provides security engineers with extra reports on hackers trying to target network devices, giving them time to study logs, analyze threats, and prepare for an attack.

\subsection{Instrumentation and Simulation}
In this paper, the goal is to develop a secure network topology with two different Honeypots, each serving a different purpose. We used the GNS3 simulator for our lab to visualize the project in a virtual environment \cite{kurniawan2019implementation}.

The Graphical Network Simulator-3 (GNS3) is a free, open-source network simulation platform. It provides a graphical interface where complex network labs can be created by importing various network equipment like routers, switches, firewalls, and servers. It works by emulating the images of the devices that do the actual job. Using this emulator, anyone can design a high-quality, complex network topology, including simulations of Ethernet, frame relay, and ATM switches. Virtual machines from VMWare or VirtualBox can be imported using the GNS3 VM. GNS3 has a built-in feature for packet capture using Wireshark. The main drawback of GNS3 is that while the platform itself doesn't use many base machine CPU resources, the lab machines (routers, switches, firewalls) installed in GNS3 do consume CPU and RAM according to their allocated memory \cite{castillo2020gns3}.

Our lab was built with seven components, including a cloud, a distribution switch, and another switch from the GNS3 simulator. We also used a Cisco IOS router, a Pentbox Honeypot, a KFSensor Honeypot, and a Firewall in our topology. The Pentbox and KFSensor honeypots were run in VMWare, and the virtual machines were imported into the GNS3 environment as VMWare templates.

\begin{figure}[ht]
    \centering
    \includegraphics[width=0.99\linewidth]{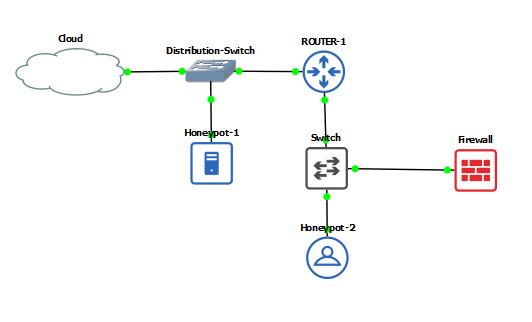}
    \caption{GNS3 Topology diagram.}
    \label{fig:gns_topology}
\end{figure}

The entire lab was run in the GNS VM, with both GNS3 and GNS VM at version 2.2.5. The GNS VM was allocated 10 GB of RAM and 2 dual-core processors. Using the GNS VM is better than the standalone GNS3, especially when running Cisco IOS or Unix devices. Once installed, the GNS VM is simple to use as a plug-and-play OS. It can also be accessed remotely as a cloud by setting up internet connectivity \cite{kazak2018methods}.

\begin{figure}[ht]
    \centering
    \includegraphics[width=0.99\linewidth]{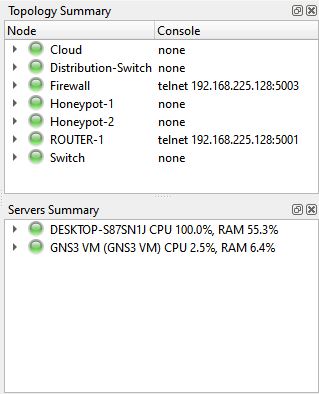}
    \caption{GNS3 topology and servers summary.}
    \label{fig:gns_summary}
\end{figure}

We ran two of our honeypots in VMWare and imported them as templates into the GNS3 VM environment. One honeypot was installed on Ubuntu, and the other on a Windows operating system.

\section{Data Collection and Analysis}
The Ubuntu-based Pentbox honeypot was used to mimic an attacker's target. We lured the attacker with a fake banner on a web server using this honeypot, causing them to leave behind a footprint. This data is useful for forensic analysis of the attacker. Pentbox is a penetration testing tool with several features for penetration testing, network analysis, and monitoring \cite{wang2021honeypots}.

The Pentbox was set up as a web server with a fake message, and port 8080 was opened to fool the attacker. This honeypot serves as the first decoy in front of the network. This open-source program can be cloned from GitHub for learning purposes \cite{ravji2018integrated}. This honeypot is a good network decoy that can gather an attacker's footprints before they can assault the main network.

\begin{figure}[ht]
    \centering
    \includegraphics[width=0.99\linewidth]{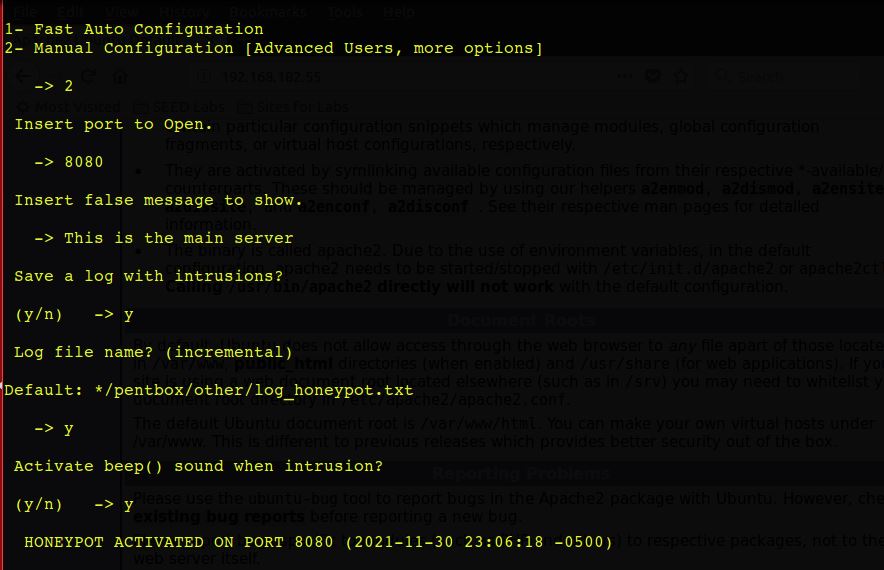}
    \caption{Pentbox configuration data as a Honeypot.}
    \label{fig:pentbox_config}
\end{figure}

The router connects and distributes the internet to all connected devices, routing IP packets between them. It translates network addresses and sends local network traffic to the wide area network and vice versa. Routers also have security features like access control lists, port filtering, and IP filtering. Although routers do not provide deep inspection services like firewalls, some can offer content filtering and bandwidth control \cite{szewczyk2017broadband}.

Our routing device handled LAN-to-WAN routing and acted as the network's border gateway. Border routers manage Border Gateway Protocol (BGP) and internet traffic. With our CISCO IOS 7200 border router, we can filter traffic from specific sources and destinations. We can also set port restrictions for security. The border router can filter IP addresses, access control lists, and ports to prevent network attacks, allowing us to block any suspicious or vulnerable addresses. We used the CISCO 7200 border router because it provides these features and serves as a reliable routing system with security surveillance capabilities \cite{khattakrole, cui2011killing}.

\begin{figure}[ht]
    \centering
    \includegraphics[width=0.99\linewidth]{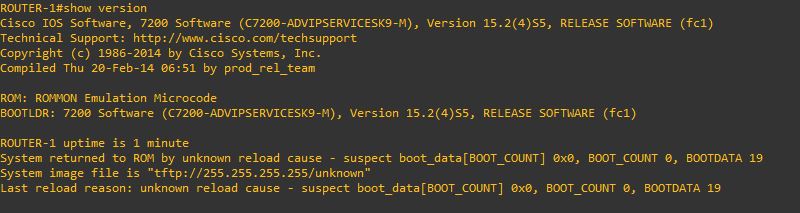}
    \caption{CISCO IOS Router version.}
    \label{fig:cisco_ios}
\end{figure}

KFSensor is an intrusion detection system for Windows that can be used as a honeypot in a network security model. It is known as an active intrusion detection tool and can identify attackers moving on a network and store logs to analyze their motives. KFSensor provides an interface where anyone can view a real-time threat report with detailed information \cite{cui2011killing}.

\begin{figure}[ht]
    \centering
    \includegraphics[width=0.99\linewidth]{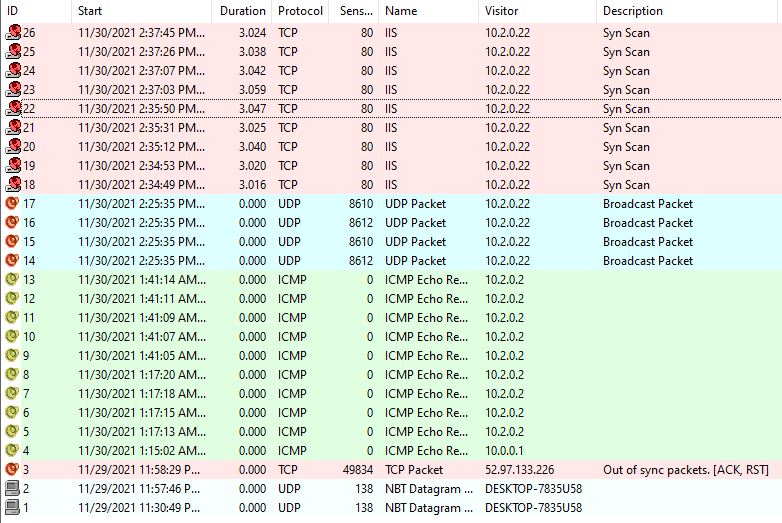}
    \caption{Analyzing intruder behavior using KFSensor.}
    \label{fig:kfsensor_behavior}
\end{figure}

In this research, we used KFSensor as a honeypot to analyze attacker behavior and attack patterns. KFSensor provides information about the attacker's source, protocol, traffic, packet type, and more \cite{naik2018honeypots}. I generated threats from a testing computer and observed the results on its real-time threat interface. This interface showed TCP, UDP, and ICMP threats from other machines. TCP and UDP packets came from the testing attacker's workstation, while ICMP packets came from the router. We found broadcast UDP packets and a 'sync scan' TCP, which is caused by a Distributed Denial of Service (DDoS) attack from another machine on the network.

In this lab, we generated a DDoS attack from an attacker workstation to analyze the data collected by KFSensor. We used Slowloris in Kali Linux for the DDoS attack. Slowloris is an open-source tool available on GitHub that generates many HTTP session requests to a target location for an HTTP DDoS attack. The targeted system becomes unable to respond to other users due to the excessive HTTP requests, causing service interruptions \cite{yevsieieva2017analysis}. Slowloris generated a DDoS attack by sending keep-alive header packets.

We sent a DDoS attack to the honeypot to test how we could collect data for analyzing the attacker's patterns. This DDoS attack drains service resources such as bandwidth, CPU, and memory, causing service interruptions \cite{shorey2018performance}. The Slowloris tool sent a large number of HTTP requests to our honeypot to verify that it captures the attacker's data in real time. We measured the attack's start and end times in the event description. The figure shows that the attack was a form of sniffing, which we measured. After reviewing multiple reports on the time frame and action type, we can determine the attacker's timing and patterns, including when they choose to launch a new attack.

When more information about the attacker is needed, we can use the KFSensor Nmap feature to trace them. We tracked the path from the attacker's IP information using Nmap and scanned it to get information about the operating system, system version, connectivity status, and packet travel time through a trace report \cite{naik2018honeypots}. With all this information, we measured the distance between our system and the attacker's system and gathered details about the materials used to generate the attack. Knowing the operating system and version helps us understand how dangerous the next attack could be.

\begin{figure}[ht]
    \centering
    \includegraphics[width=0.99\linewidth]{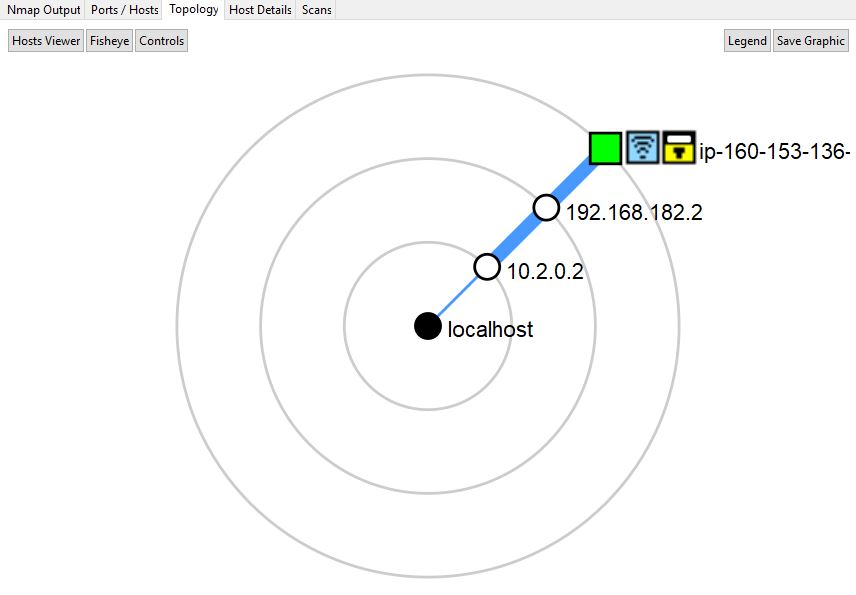}
    \caption{KFSensor Nmap topology finder.}
    \label{fig:kfsensor_nmap}
\end{figure}

To find the attacker's location on a global map, we used the Nmap topology to get the targeted IP address. The Nmap tool built into KFSensor can locate the vulnerable IP address by following the trace route step-by-step and provides graphical feedback about the attacker's hop location \cite{naik2018honeypots}. At this point, it is clear that we have successfully back-traced the attacker using the KFSensor Honeypot. We used KFSensor Nmap to scan and capture information about the attacker's system. With Nmap in KFSensor, we scanned every active IP address, performed a full network scan, identified network vulnerabilities, and developed a visual map of the attacker’s location, open ports, and other key information \cite{naik2018honeypots}.

\section{Result and Performance Analysis}
The designed network security diagram set a benchmark, and we tested its performance. The goal was to reduce network vulnerability and challenge attackers by analyzing and observing the collected reports. Our first approach was to create a fake environment to trap the attacker. The first honeypot we set up was designed to mimic a fake web server on port 8080 to trick the intruder. The server would display a fake message upon entry.

The second honeypot was designed to attract the intruder to get their footprints and analyze scan reports. This device provides a brief report that allows us to identify and observe the intruder's attitude and attack patterns.

The performance was analyzed in a real-time attack scenario. We connected all devices to the internet to make them reachable globally and observed how attackers were more likely to find and attack our devices.

The result was that the first honeypot received the highest number of vulnerability scanning reports. This is because this honeypot is at the front of the network, and its information can be easily found by scanning the network surface. We analyzed this report using the built-in Wireshark on the Ubuntu operating system of that honeypot. Wireshark is a network vulnerability analysis tool available on most popular operating systems, and it can capture packets and traffic patterns going in and out of a specific machine.

\begin{figure}[ht]
    \centering
    \includegraphics[width=0.99\linewidth]{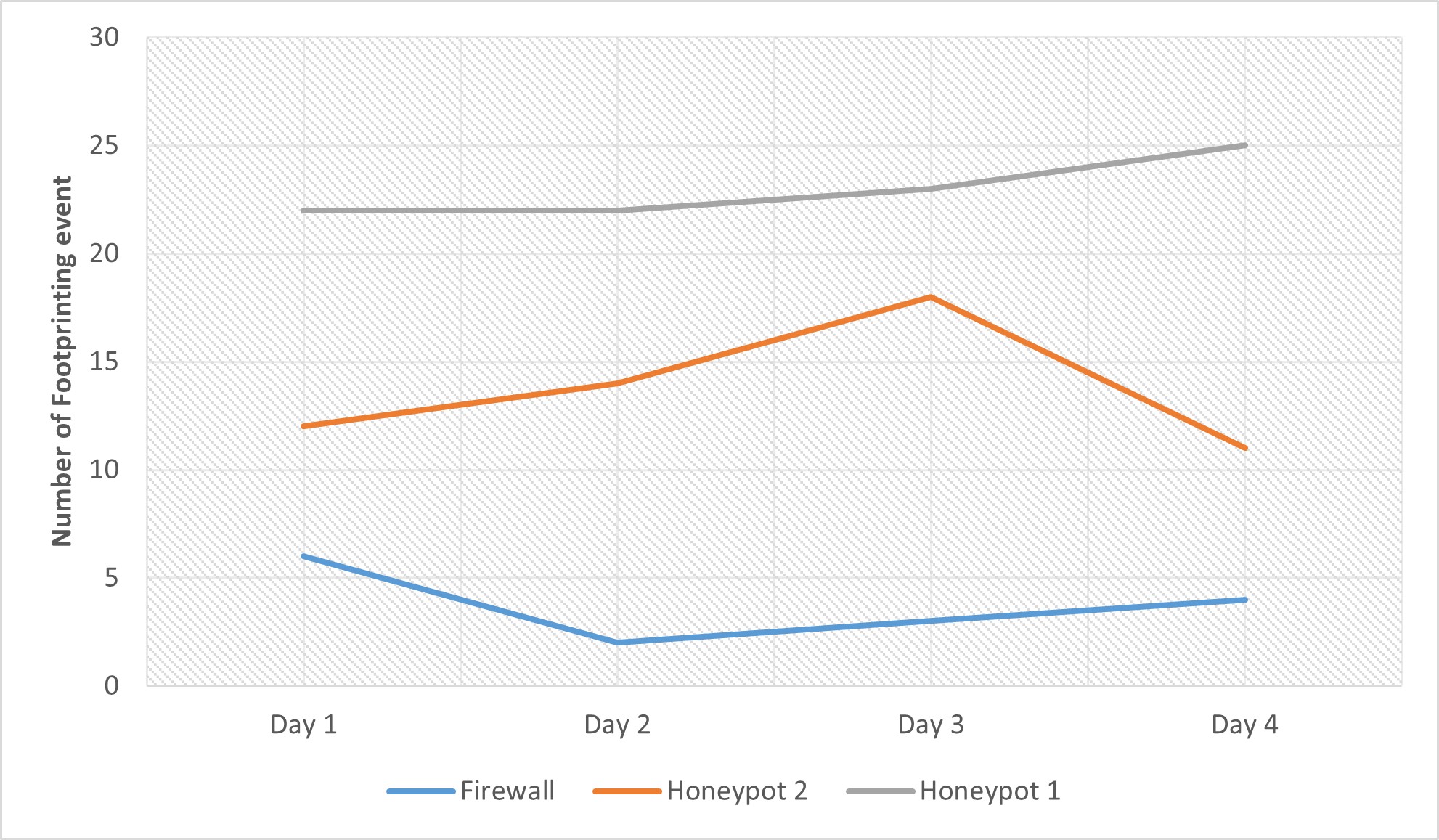}
    \caption{Footprint events on Honeynet devices and Firewall.}
    \label{fig:footprint_events}
\end{figure}

We observed that Honeypot 2, which was used for analyzing intruders, received fewer hits from footprints and vulnerability scans from the internet. This was due to the Honeynet router, which has its own defense system against flooding and Denial of Service attacks to keep the routing engine running smoothly. Even though we did not set up any access control lists or port filtering, the router has a built-in protection system to prevent attacks from taking down its core operations.

\begin{figure}[ht]
    \centering
    \includegraphics[width=0.99\linewidth]{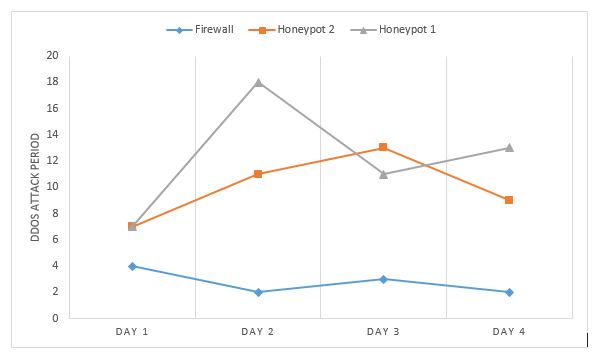}
    \caption{DDoS events on Honeynet devices and Firewall.}
    \label{fig:ddos_events}
\end{figure}

While connected to the global internet, we discovered another major attack on our devices: a Distributed Denial of Service (DDoS) attack. Similar to the footprinting events, we found that Honeynet-1 received most of the attack events, while the Honeynet and the Firewall received the fewest. This attack caused higher bandwidth consumption and CPU usage on each device. The firewall has its own mechanism to protect against DoS attacks. In general, DDoS attacks are carried out by botnets that send a huge amount of traffic or service requests to the target, causing service interruptions by overwhelming the system's processes.

\begin{table}[ht]
    \centering
    \caption{Result of foot printing on each device from WAN.}
    \begin{tabular}{|c|c|c|c|}
        \hline
        & Firewall & Honeypot 2 & Honeypot 1 \\
        \hline
        Day 1 & 6 & 12 & 22 \\
        \hline
        Day 2 & 2 & 14 & 22 \\
        \hline
        Day 3 & 3 & 18 & 23 \\
        \hline
        Day 4 & 4 & 11 & 25 \\
        \hline
    \end{tabular}
    \label{tab:footprinting}
\end{table}

By observing the scan and footprint data, it is clear that the first honeypot attracted the most attention from attackers, while the second honeypot received less, as it was inside the Honeynet. The Firewall attracted the least attention because attackers had already found access to two other devices before reaching it.

\begin{table}[h]
    \centering
    \caption{Report of Distributed Denial of Service attacks on each device from WAN.}
    \begin{tabular}{|c|c|c|c|}
        \hline
        & Honeypot 1 & Honeypot 2 & Firewall \\
        \hline
        Day 1 & 7 & 7 & 4 \\
        \hline
        Day 2 & 18 & 11 & 2 \\
        \hline
        Day 3 & 11 & 13 & 3 \\
        \hline
        Day 4 & 13 & 9 & 2 \\
        \hline
    \end{tabular}
    \label{tab:ddos_report}
\end{table}

The DDoS attack results were similar, suggesting that attackers performed some footprinting on the network before launching the attack. Generally, attackers gather information before an attack, searching for opportunities and weak points. This is the first phase of hacking any system.

It is likely that attackers spent more time on the honeypots, which is why they invested less time on the firewall. This gave security engineers enough time to analyze the risk by studying the data from the honeypots. The next step would be to apply filters or create limitations on the firewall. Keeping an eye on the attacker’s source and the ports they attack can also help security engineers prepare.

The goal of this work was to mislead attackers with a decoy and analyze their behavior before an attack. The results demonstrated the project's success, as most of the major attacks were directed at the honeypots, and hackers remained on the Honeynet for a long time instead of the real infrastructure. Additionally, we received a detailed report on their attack methods, which helped us decide on the next stage of security. The hybrid security model showed that these decoys gave security engineers enough time to prepare before any security breach or service interruption. By tracing these reports, any security engineer can easily determine the next steps for improving defense. Tracing the attacker’s steps is highly recommended for protecting a network before experiencing a cyber attack.

\section{Future Work}
 Honeypots and Honeynets are among the most powerful tools for security research, and this study describes some aspects of a security project using them. However, the author intends to continue researching network security and a hybrid network architecture model to improve network infrastructure and security. Further study on this paper may include the following topics. The honeypots should be enhanced to resemble a real server with several services running in different VMs. The Honeynet will use a clustered, load-balanced network design with multiple routers for different internet service providers, merging internet traffic effectively. We will move beyond DDoS attacks to integrate intrusion protection into the honeypot along with an intrusion detection system. The Honeynet infrastructure will be automated to integrate with the main network infrastructure. This study discusses Honeynet and Honeypot solutions with practical examples. The author plans to add more sophisticated network scenarios and honeypot servers in future work.

\section{Conclusion}
Network security plays a vital role in protecting confidential data and important services from theft and can save any company from cyber robbery. By focusing on this subject, this paper provided a practical concept with a real-life attack scenario on network security. The solution showed how a honeypot can be a strong security tool by providing different services. When we need to misguide a hacker or collect their footprints from a safe system, a honeypot is undoubtedly one of the best solutions. The Honeynet was presented as one of the most effective tools to protect a network, supplementing firewall security by using honeypot devices. Overall, this project discussed the importance of network security with a technique for implementing Honeypot and Honeynet services in a production network.

\bibliographystyle{IEEEtran}

\bibliography{references}

@article{rathore2017social,
  title={Social network security: Issues, challenges, threats, and solutions},
  author={Rathore, Shailendra and Sharma, Pradip Kumar and Loia, Vincenzo and Jeong, Young-Sik and Park, Jong Hyuk},
  journal={Information sciences},
  volume={421},
  pages={43--69},
  year={2017},
  publisher={Elsevier}
}

@article{lin2018survey,
  title={A survey on network security-related data collection technologies},
  author={Lin, Huaqing and Yan, Zheng and Chen, Yu and Zhang, Lifang},
  journal={IEEe Access},
  volume={6},
  pages={18345--18365},
  year={2018},
  publisher={IEEE}
}

@inproceedings{sagduyu2019iot,
  title={IoT network security from the perspective of adversarial deep learning},
  author={Sagduyu, Yalin E and Shi, Yi and Erpek, Tugba},
  booktitle={2019 16th Annual IEEE International Conference on Sensing, Communication, and Networking (SECON)},
  pages={1--9},
  year={2019},
  organization={IEEE}
}

@inproceedings{mozhaev2017multiservice,
  title={Multiservice network security metric},
  author={Mozhaev, Oleksandr and Kuchuk, Heorgii and Kuchuk, Nina and Mozhaev, Mykhailo and Lohvynenko, Mykhailo},
  booktitle={2017 2nd International Conference on Advanced Information and Communication Technologies (AICT)},
  pages={133--136},
  year={2017},
  organization={IEEE}
}

@inproceedings{matin2019malware,
  title={Malware detection using honeypot and machine learning},
  author={Matin, Iik Muhamad Malik and Rahardjo, Budi},
  booktitle={2019 7th international conference on cyber and IT service management (CITSM)},
  volume={7},
  pages={1--4},
  year={2019},
  organization={IEEE}
}

@article{ren2021theoretical,
  title={A theoretical method to evaluate honeynet potency},
  author={Ren, Jianguo and Zhang, Chunming and Hao, Qihong},
  journal={Future Generation Computer Systems},
  volume={116},
  pages={76--85},
  year={2021},
  publisher={Elsevier}
}

@inproceedings{fox2019deployment,
  title={The deployment of an IoT network infrastructure, as a localised regional service},
  author={Fox, John and Donnellan, Andrew and Doumen, Liam},
  booktitle={2019 IEEE 5th World Forum on Internet of Things (WF-IoT)},
  pages={319--324},
  year={2019},
  organization={IEEE}
}

@article{sheikh2020network,
  title={Network Fundamentals and Infrastructure Security},
  author={Sheikh, Ahmed F and Sheikh, Ahmed F},
  journal={CompTIA Security+ Certification Study Guide: Network Security Essentials},
  pages={9--34},
  year={2020},
  publisher={Springer}
}

@article{baykara2018novel,
  title={A novel honeypot based security approach for real-time intrusion detection and prevention systems},
  author={Baykara, Muhammet and Das, Resul},
  journal={Journal of Information Security and Applications},
  volume={41},
  pages={103--116},
  year={2018},
  publisher={Elsevier}
}

@article{krishnaveni2018survey,
  title={A survey on honeypot and honeynet systems for intrusion detection in cloud environment},
  author={Krishnaveni, S and Prabakaran, S and Sivamohan, S},
  journal={Journal of Computational and Theoretical Nanoscience},
  volume={15},
  number={9-10},
  pages={2949--2953},
  year={2018},
  publisher={American Scientific Publishers}
}

@article{leech2024ten,
  title={Ten hard problems in artificial intelligence we must get right},
  author={Leech, Gavin and Garfinkel, Simson and Yagudin, Misha and Briand, Alexander and Zhuravlev, Aleksandr},
  journal={arXiv preprint arXiv:2402.04464},
  year={2024}
}

@article{malecki2019best,
  title={Best practices for preventing and recovering from a ransomware attack},
  author={Malecki, Florian},
  journal={Computer Fraud \& Security},
  volume={2019},
  number={3},
  pages={8--10},
  year={2019},
  publisher={MA Business London}
}

@inproceedings{kyriakou2018container,
  title={Container-based honeypot deployment for the analysis of malicious activity},
  author={Kyriakou, Andronikos and Sklavos, Nicolas},
  booktitle={2018 Global Information Infrastructure and Networking Symposium (GIIS)},
  pages={1--4},
  year={2018},
  organization={IEEE}
}

@article{fan2019honeydoc,
  title={Honeydoc: an efficient honeypot architecture enabling all-round design},
  author={Fan, Wenjun and Du, Zhihui and Smith-Creasey, Max and Fernandez, David},
  journal={IEEE journal on selected areas in communications},
  volume={37},
  number={3},
  pages={683--697},
  year={2019},
  publisher={IEEE}
}

@inproceedings{dowling2017zigbee,
  title={A ZigBee honeypot to assess IoT cyberattack behaviour},
  author={Dowling, Seamus and Schukat, Michael and Melvin, Hugh},
  booktitle={2017 28th Irish signals and systems conference (ISSC)},
  pages={1--6},
  year={2017},
  organization={IEEE}
}

@article{wu2019active,
  title={Active defense-based resilient sliding mode control under denial-of-service attacks},
  author={Wu, Chengwei and Wu, Ligang and Liu, Jianxing and Jiang, Zhong-Ping},
  journal={IEEE Transactions on Information Forensics and Security},
  volume={15},
  pages={237--249},
  year={2019},
  publisher={IEEE}
}

@article{aminzade2018confidentiality,
  title={Confidentiality, integrity and availability--finding a balanced IT framework},
  author={Aminzade, Michael},
  journal={Network Security},
  volume={2018},
  number={5},
  pages={9--11},
  year={2018},
  publisher={MA Business London}
}

@article{yin2020hierarchically,
  title={Hierarchically defining Internet of Things security: From CIA to CACA},
  author={Yin, Lihua and Fang, Binxing and Guo, Yunchuan and Sun, Zhe and Tian, Zhihong},
  journal={International Journal of Distributed Sensor Networks},
  volume={16},
  number={1},
  pages={1550147719899374},
  year={2020},
  publisher={SAGE Publications Sage UK: London, England}
}

@article{atieh2021assuring,
  title={Assuring the optimum security level for network, physical and cloud infrastructure},
  author={Atieh, Ali T},
  journal={ScienceOpen Preprints},
  year={2021},
  publisher={ScienceOpen}
}

@inproceedings{saigushev2018information,
  title={Information systems at enterprise. Design of secure network of enterprise},
  author={Saigushev, NY and Mikhailova, UV and Vedeneeva, OA and Tsaran, AA},
  booktitle={Journal of Physics: Conference Series},
  volume={1015},
  number={4},
  pages={042054},
  year={2018},
  organization={IOP Publishing}
}

@article{nur2018record,
  title={Record route IP traceback: Combating DoS attacks and the variants},
  author={Nur, Abdullah Yasin and Tozal, Mehmet Engin},
  journal={Computers \& Security},
  volume={72},
  pages={13--25},
  year={2018},
  publisher={Elsevier}
}

@inproceedings{tirumala2019survey,
  title={A survey on cybersecurity awareness concerns, practices and conceptual measures},
  author={Tirumala, SS and Valluri, Maheswara Rao and Babu, GA},
  booktitle={2019 International Conference on Computer Communication and Informatics (ICCCI)},
  pages={1--6},
  year={2019},
  organization={IEEE}
}

@inproceedings{jingyao2019securing,
  title={Securing a network: how effective using firewalls and VPNs are?},
  author={Jingyao, Sun and Chandel, Sonali and Yunnan, Yu and Jingji, Zang and Zhipeng, Zhang},
  booktitle={Future of Information and Communication Conference},
  pages={1050--1068},
  year={2019},
  organization={Springer}
}

@inproceedings{wildenauer2019hacking,
  title={Hacking an optics manufacturing machine: You don't see it coming?!},
  author={Wildenauer, Robert and Leidl, Karl and Schramm, Martin},
  booktitle={Sixth European Seminar on Precision Optics Manufacturing},
  volume={11171},
  pages={35--40},
  year={2019},
  organization={SPIE}
}

@inproceedings{kurniawan2019implementation,
  title={Implementation and analysis ipsec-vpn on cisco asa firewall using gns3 network simulator},
  author={Kurniawan, Dwi Ely and Arif, Hamdani and Nelmiawati, N and Tohari, Ahmad Hamim and Fani, Maidel},
  booktitle={Journal of Physics: Conference Series},
  volume={1175},
  number={1},
  pages={012031},
  year={2019},
  organization={IOP Publishing}
}

@inproceedings{castillo2020gns3,
  title={GNS3 limitations when emulating connectivity and management for backbone networks: a case study of CANARIE},
  author={Castillo-Vel{\'a}zquez, Jose-Ignacio and Delgado-Villegas, Alonso},
  booktitle={2020 IEEE Canadian Conference on Electrical and Computer Engineering (CCECE)},
  pages={1--4},
  year={2020},
  organization={IEEE}
}

@inproceedings{kazak2018methods,
  title={Methods and Tools for Evaluating the Accuracy of the Air Navigation Using GNS},
  author={Kazak, VM and Shevchuk, DO and Panchuk, LV and Shulevka, VV},
  booktitle={2018 IEEE 5th International Conference on Methods and Systems of Navigation and Motion Control (MSNMC)},
  pages={79--82},
  year={2018},
  organization={IEEE}
}

@article{wang2021honeypots,
  title={Honeypots and knowledge for network defense},
  author={Wang, Ping and D’Cruze, Hubert},
  journal={Issues in Information Systems},
  volume={22},
  number={3},
  pages={241--254},
  year={2021}
}

@inproceedings{ravji2018integrated,
  title={Integrated intrusion detection and prevention system with honeypot in cloud computing},
  author={Ravji, Sajaan and Ali, Maaruf},
  booktitle={2018 International Conference on Computing, Electronics \& Communications Engineering (iCCECE)},
  pages={95--100},
  year={2018},
  organization={IEEE}
}

@article{szewczyk2017broadband,
  title={Broadband router security: History, challenges and future implications},
  author={Szewczyk, Patryk and Macdonald, Rose},
  year={2017},
  publisher={Association of Digital Forensics, Security and Law}
}

@inproceedings{khattakrole,
  title={Role of Border Router in 6LoWPAN Security},
  author={Khattak, Fida and Ginzboorg, Philip and Niemi, Valtteri and Ekberg, Jan-Erik},
  booktitle={Workshop on Smart Object Security}
}

@inproceedings{cui2011killing,
  title={Killing the Myth of Cisco $\{$IOS$\}$ Diversity: Recent Advances in Reliable Shellcode Design},
  author={Cui, Ang and Kataria, Jatin and Stolfo, Salvatore J},
  booktitle={5th USENIX Workshop on Offensive Technologies (WOOT 11)},
  year={2011}
}

@inproceedings{naik2018honeypots,
  title={Honeypots that bite back: A fuzzy technique for identifying and inhibiting fingerprinting attacks on low interaction honeypots},
  author={Naik, Nitin and Jenkins, Paul and Cooke, Roger and Yang, Longzhi},
  booktitle={2018 IEEE International Conference on fuzzy systems (FUZZ-IEEE)},
  pages={1--8},
  year={2018},
  organization={IEEE}
}

@inproceedings{shorey2018performance,
  title={Performance comparison and analysis of slowloris, goldeneye and xerxes ddos attack tools},
  author={Shorey, Tanishka and Subbaiah, Deepthi and Goyal, Ashwin and Sakxena, Anuraag and Mishra, Alekha Kumar},
  booktitle={2018 International Conference on Advances in Computing, Communications and Informatics (ICACCI)},
  pages={318--322},
  year={2018},
  organization={IEEE}
}

\end{document}